\documentclass[a4paper]{article}
\pdfoutput=1

\usepackage[T1]{fontenc}
\usepackage{authblk}
\usepackage{amsmath}
\usepackage{textcomp}% for upright mu (\textmu)
\usepackage{graphicx}
\usepackage{url}
\usepackage[numbers,sort&compress]{natbib}
\newcommand{\wraprefprepost}[3]{\wrapprepost{#1}{#2}{\ref{#3}}}
\newcommand{\formatrefplain}{\ref}
\newcommand{\formatrefparens}{\wraprefprepost{(}{)}}

\newcommand{\wrapprepost}[3]{{#1}{#3}{#2}}

\newcommand{\tagwithlabel}[2]{#1~#2}

\newcommand{\makelabeledcrossrefmacro}[4]
	{\newcommand{#3}{#1{#4}{#2}}}
\newcommand{\makecrossrefmaker}[3]
	{\newcommand{#1}{\makelabeledcrossrefmacro{#2}{#3}}}

\newcommand{\eqnrefformat}{\formatrefparens}
\newcommand{\eqnlabelbinding}{\tagwithlabel}
\newcommand{\eqnlabel}{eq.}
\newcommand{\Eqnlabel}{Eq.}
\newcommand{\eqnslabel}{eqs.}
\newcommand{\Eqnslabel}{Eqs.}

\newcommand{\eqnnum}{\eqnrefformat}
\makecrossrefmaker{\newlabeledeqnref}{\eqnlabelbinding}{\eqnnum}
\makecrossrefmaker{\newwordpluseqnref}{\tagwithlabel}{\eqnnum}

\newlabeledeqnref{\eqn}{\eqnlabel}
\newlabeledeqnref{\Eqn}{\Eqnlabel}
\newlabeledeqnref{\eqns}{\eqnslabel}
\newlabeledeqnref{\Eqns}{\Eqnslabel}

\newwordpluseqnref{\andeqn}{and}
\newwordpluseqnref{\througheqn}{through}

\newcommand{\figrefformat}{\formatrefplain}
\newcommand{\figlabelbinding}{\tagwithlabel}
\newcommand{\figlabel}{fig.}
\newcommand{\Figlabel}{Fig.}
\newcommand{\figslabel}{figs.}
\newcommand{\Figslabel}{Figs.}

\newcommand{\fignum}{\figrefformat}
\makecrossrefmaker{\newlabeledfigref}{\figlabelbinding}{\fignum}
\makecrossrefmaker{\newwordplusfigref}{\tagwithlabel}{\fignum}

\newlabeledfigref{\fig}{\figlabel}
\newlabeledfigref{\Fig}{\Figlabel}
\newlabeledfigref{\figs}{\figslabel}
\newlabeledfigref{\Figs}{\Figslabel}

\newwordplusfigref{\andfig}{and}
\newwordplusfigref{\throughfig}{through}

\newcommand{\sxnrefformat}{\formatrefplain}
\newcommand{\sxnlabelbinding}{\tagwithlabel}
\newcommand{\sxnlabel}{section}
\newcommand{\Sxnlabel}{Section}
\newcommand{\sxnslabel}{sections}
\newcommand{\Sxnslabel}{Sections}

\newcommand{\sxnnum}{\sxnrefformat}
\makecrossrefmaker{\newlabeledsxnref}{\sxnlabelbinding}{\sxnnum}
\makecrossrefmaker{\newwordplussxnref}{\tagwithlabel}{\sxnnum}

\newlabeledsxnref{\sxn}{\sxnlabel}
\newlabeledsxnref{\Sxn}{\Sxnlabel}
\newlabeledsxnref{\sxns}{\sxnslabel}
\newlabeledsxnref{\Sxns}{\Sxnslabel}
	
\newwordplussxnref{\andsxn}{and}
\newwordplussxnref{\throughsxn}{through}

\newcommand{\tblrefformat}{\formatrefplain}
\newcommand{\tbllabelbinding}{\tagwithlabel}
\newcommand{\tbllabel}{table}
\newcommand{\Tbllabel}{Table}
\newcommand{\tblslabel}{tables}
\newcommand{\Tblslabel}{Tables}

\newcommand{\tblnum}{\tblrefformat}
\makecrossrefmaker{\newlabeledtblref}{\tbllabelbinding}{\tblnum}
\makecrossrefmaker{\newwordplustblref}{\tagwithlabel}{\tblnum}

\newlabeledtblref{\tbl}{\tbllabel}
\newlabeledtblref{\Tbl}{\Tbllabel}
\newlabeledtblref{\tbls}{\tblslabel}
\newlabeledtblref{\Tbls}{\Tblslabel}
	
\newwordplustblref{\andtbl}{and}
\newwordplustblref{\throughtbl}{through}

\usepackage{xspace}

\newcommand{\ie}{i.e.}
\newcommand{\etc}{etc.} 
\newcommand{\eg}{e.g.}
\newcommand{\brim}{BRIM\xspace}
\newcommand{\RD}{R\&{}D\xspace}

\newcommand{\foreign}{\textit}
\newcommand{\newterm}[1]{#1}
\newcommand{\subjectindex}[1]{\textsl{#1}}
\newcommand{\tablehead}{\multicolumn}
\newcommand{\missingtableentry}{\multicolumn{1}{c}{---}}

\newcommand{\acknowledgments}
	{\section*{Acknowledgments}\label{sec:acknowledgments}} 

\newcommand{\largeparens}[1]{\left ( #1 \right )}
\newcommand{\largesquare}[1]{\left [ #1 \right ]}

\newcommand{\largestraight}[1]{\left | #1 \right |}

\newcommand{\by}{\times}
\newcommand{\given}{\mid}

\newcommand{\abs}{\largestraight}
\newcommand{\logit}[1]{\mathit{Logit}\largeparens{#1}}
\newcommand{\loglikelihood}[1]{\log L\largeparens{#1}}

\newcommand{\mathpuncspace}{\quad}
\newcommand{\mathcomma}{\mathpuncspace,}
\newcommand{\mathperiod}{\mathpuncspace.}

\newcommand{\mat}[1]{\boldsymbol{#1}}
\newcommand{\matelem}[3]{#1_{#2#3}}
\newcommand{\vctr}[1]{\boldsymbol{#1}}
\newcommand{\vctrelem}[2]{#1_{#2}}

\newcommand{\litvctr}[1]{\largeparens{#1}}

\newcommand{\matrixdims}[2]{\ensuremath{#1 \by #2}}
\newcommand{\transpose}[1]{#1^\mathrm{T}}

\newcommand{\zeromat}{\mat{O}}

\newcommand{\expectationop}[1]{E\largesquare{#1}}
\newcommand{\varianceop}[1]{\mathit{Var}\largesquare{#1}}
\newcommand{\probfunc}[1]{\mathrm{Pr}\largeparens{#1}}

\newcommand{\vertexdegree}[1]{\ensuremath{d_{#1}}}

\newcommand{\adjmat}{\mat{A}}
\newcommand{\adjelem}[2]{A_{#1#2}}
\newcommand{\adjsubmat}{\mat{M}}

\newcommand{\probelem}[2]{P_{#1#2}}

\newcommand{\numvertices}{\ensuremath{N}}
\newcommand{\numedges}{\ensuremath{M}}
\newcommand{\modularity}{\ensuremath{Q}}

\newcommand{\freq}[1]{f\largeparens{#1}}
\newcommand{\topfreq}[2]{\topfreqnoarg{#1}\largeparens{#2}}
\newcommand{\topfreqnoarg}[1]{f_{#1}}

\newcommand{\matrixbrackets}{\largesquare}

\newcommand{\bipartsubstructure}[1]{\matrixbrackets{
	\begin{array}{cc}
		\zeromat & #1 \\
		\transpose{#1} & \zeromat
	\end{array}}}

\newcommand{\topicmetric}[1]{d_{#1}}

\newcommand{\orgpair}[2]{(\( #1 \), \( #2 \))}

\newcommand{\numexpvars}{\ensuremath{K}}

\newcommand{\collabmat}{\mat{Y}}
\newcommand{\collabelem}[2]{\matelem{Y}{#1}{#2}}

\newcommand{\hiddenprofitelem}[2]{\matelem{Y}{#1}{#2}^{*}}

\newcommand{\oneprob}[2]{\pi_{#1#2}}
\newcommand{\zeroprob}[2]{1-\oneprob{#1}{#2}}

\newcommand{\collabmeanelem}[2]{\matelem{\mu}{#1}{#2}}
\newcommand{\collabvarelem}[2]{\matelem{\sigma^{2}}{#1}{#2}}

\newcommand{\covariatemat}[1]{\mat{X}^{(#1)}}
\newcommand{\covariateelem}[3]{\matelem{X^{(#1)}}{#2}{#3}}

\newcommand{\paramwtvctr}{\vctr{\beta}}
\newcommand{\paramwtelem}[1]{\vctrelem{\beta}{#1}}
\newcommand{\paramestvctr}{\vctr{\hat{\beta}}}

\newcommand{\covwtsumelem}[2]{\matelem{h}{#1}{#2}}

\newcommand{\variancematop}[1]{\mat{V}\largeparens{#1}}

\newcommand{\obscovelem}[3]{\matelem{x^{(#1)}}{#2}{#3}}
\newcommand{\specvctr}[1]{\vctr{s}_{#1}}
\newcommand{\numprojpart}[2]{N_{#1,#2}}
\newcommand{\totprojpart}[1]{N_{#1}}
\newcommand{\numprojknown}[1]{P_{i}}

\newcommand{\numshortestbetthru}[3]{B\largeparens{#1,#2;#3}}
\newcommand{\numshortestbet}[2]{B\largeparens{#1,#2}}
\newcommand{\betcent}[1]{b\largeparens{#1}}

\newcommand{\clustcoef}{CC_{1}}
\newcommand{\locclustcoef}[1]{\clustcoef\largeparens{#1}}
\newcommand{\degree}[1]{k_{#1}}
\newcommand{\numtris}[1]{T_{#1}}

\newcommand{\highsig}{\(^{**}\)}
\newcommand{\vhighsig}{\(^{***}\)}

\graphicspath{{pdffigs/}{epsfigs/}{mpsfigs/}}

\begin{document}

\title{Analyzing and modeling European \RD collaborations: Challenges and opportunities from a large social network} 

% \chapterauthor[Michael J. Barber,
% Manfred Paier,
% and Thomas Scherngell]
% {Michael J. Barber,
% Manfred Paier,
% and Thomas Scherngell \\ 
% Austrian Research Centers---ARC,
% Division systems research,
% Donau-City-Stra{\ss}e 1,
% 1220 Vienna, 
% Austria}

\author{Michael J. Barber}
\author{Manfred Paier}
\author{Thomas Scherngell}
\affil{Austrian Research Centers---ARC, 
Division systems research,
Donau-City-Stra{\ss}e 1,
1220 Vienna, 
Austria}

\maketitle

\section{Introduction} \label{sec:intro}

Networks have attracted a burst of attention in the last decade
(useful reviews include
references~\cite{ChrAlb:2007,DorMen:2004,New:2003,AlbBar:2002}), with
applications to natural, social, and technological systems. While
networks provide a powerful abstraction for investigating relationships and interactions,
the preparation and analysis of complex real-world networks nonetheless
presents significant challenges. 
In particular social networks are characterized by a  large number of different properties 
and generation mechanisms which require a rich set of indicators. 
The objective of the current study is to analyze large social networks with respect to their 
community structure and mechanisms of network formation.  As a case study, we consider 
networks derived from the European Union's  Framework Programs (FPs) for Research and Technological Development. 

The EU FPs were implemented to follow two main strategic objectives:
First, strengthening the scientific and technological bases of
European industry to foster international competitiveness and,
second, the promotion of research activities in support of other
EU policies. In spite of their different scopes, the
fundamental rationale of the FPs has remained unchanged. All FPs
share a few common structural key elements. First, only projects
of limited duration that mobilize private and public funds at the
national level are funded. Second, the focus of funding is on
multinational and multi-actor collaborations that add value by
operating at the European level. Third, project proposals are to
be submitted by self-organized consortia and the selection for
funding is based on specific scientific excellence and socio-economic
relevance criteria \cite{RoeBar:2006b}. By considering the constituents 
of these consortia, we can represent and analyze the FPs as networks of
projects and organizations. The resulting networks are of substantial 
size, including over 50 thousand projects and over 30 thousand 
organizations.

We have general interest in studying a real-world network of large
size and high complexity from a methodological point of view.
Furthermore,  socio-economic research emphasizes the central
importance of collaborative activities in \RD for economic
competitiveness (see, for instance, reference~\cite{FagMowNel:2005}, 
among many others). Mainly for reasons of data availability,
attempts to evaluate quantitatively the structure and function of
the large social networks generated in the EU FPs have begun only
in the last few years, using social network analysis and complex
networks methodologies \cite{AlmOliLopSanMen:2007,BarFarStrStr:2008,BarKruKruRoe:2006,BreCus:2004,RoeBar:2008}. Studies to date point to the
presence of a dense and hierarchical network. A highly connected
core of frequent participants, taking leading roles within consortia,
is linked to a large number of peripheral actors, forming a giant
component that exhibits the characteristics of a small world.  

We
augment the earlier studies by applying a battery of methods to the
most recent data.  We begin with constructing the network, discussing needed processing of the raw data in \sxn{sec:dataprep} and continuing with the network definition in \sxn{sec:defnetworks}. We next examine the overall network structure in \sxn{sec:netstructure},
showing that the networks for each FP feature a giant component
with highly skewed degree distribution and small world properties.
We follow this with an exploration of community structure in \sxns{sec:commstruct} 
\andsxn{sec:fpcommunities}, showing that the networks
are made of heterogeneous subcommunities with strong topical
differentiation. Finally, we investigate determinants of network
formation with a binary choice model in \sxn{sec:binchoicemodel}; this is similar to a recent
analysis of Spanish firms \cite{ArrArr:2008}, but with a focus on the European
level and on geographic and network effects. Results are summarized in \sxn{sec:summary}.

\section{Data Preparation} \label{sec:dataprep}

We draw on the latest version of the sysres EUPRO database. This
database includes all information publicly available through the
CORDIS projects database\footnote{\url{http://cordis.europa.eu}}
and is maintained by ARC systems research (ARC sys). The sysres EUPRO database presently comprises data on funded research
projects of the EU FPs (complete for FP1--FP5, and about 70\% complete for
FP6) and all participating organizations. It contains systematic
information on project objectives and achievements, project costs,
project funding and contract type, as well as information on the participating
organizations including the full name, the full address and the
type of the organization. 

For purposes
of network analyses, the main challenge is the inconsistency of the
raw data. Apart from incoherent spelling in up to four languages
per country, organizations are labelled inhomogeneously. Entries
may range from large corporate groupings, such as EADS, Siemens and
Philips, or large public research organizations, like CNR, CNRS and
CSIC, to individual departments and labs.

Due to these shortcomings, the raw data is of limited use for meaningful
network analyses. Further, any fully automated standardization procedure is 
infeasible. Instead, a labor-intensive, manual data-cleaning process is used in building 
the database. The data-cleaning process is described in reference~\cite{RoeBar:2008}; here, 
we restrict discussion to the steps of the process relevant to the present work. These are:
\begin{enumerate}
	\item Identification of unique organization name. Organizational
	boundaries are defined by legal control. Entries are assigned
	to appropriate organizations using the more recently available
	organization name. Most records are easily identified, but,
	especially for firms, organization names may have  changed
	frequently due to mergers, acquisitions, and divestitures.
	
	\item Creation of  subentities. This is the key step for
	mitigating the bias that arises from the different scales
	at which participants appear in the data set. Ideally, we use 
	the actual group or organizational unit that
	participates in each project, but this information is only
	available for a subset of records, particularly in the case
	of firms. Instead, subentities that operate in fairly coherent activity areas
	are pragmatically defined. 
	Wherever possible, subentities are identified at the second
	lowest hierarchical tier, with each subentity
	comprising one further hierarchical sub-layer. Thus,
	universities are broken down into faculties/schools,
	consisting of departments; research organizations are broken
	down into institutes, activity areas, \etc, consisting of
	departments, groups or laboratories; and conglomerate firms
	are broken down into divisions, subsidiaries, \etc{} Subentities
	can frequently be identified from the contact information
	even in the absence of information on the actual participating
	organizational unit. 
	Note that subentities may still vary considerably in scale.
	
	\item Regionalization. The data set has been regionalized
	according to the European Nomenclature of Territorial Units
	for Statistics (NUTS) classification system\footnote{NUTS is a 
	hierarchical system of regions used by the
	statistical office of the European Community for the production of
	regional statistics. At the top of the hierarchy are NUTS-0 regions
	(countries) below which are NUTS-1 regions and then NUTS-2 regions, \etc}, where possible 
	to the NUTS3 level. Mostly, this
	has been done via information on postal codes.

\end{enumerate}
Due to resource limitations, only 
organizations appearing more than thirty times in the standardization table for FP1--FP5 have 
thus far been processed. This could bias the results; however, the networks have a structure  
such that the size of the bias is quite low (see reference~\cite{RoeBar:2008}). 

Additionally, we make use of  a representative survey\footnote{
This survey was conducted in 2007 by the Austrian Research Centers
GmbH, Vienna, Austria and operated by b-wise GmbH, Karlsruhe,
Germany. } of FP5 participants\footnote{We chose FP5 (1998-2002)
for the survey, in order to cover some of the developments over
time, including prior as well as subsequent bilateral collaborations,
and effects of the collaboration both with respect to scientific
and commercial outcome. Thus, the survey is able to complement the
sysres EUPRO database.}.
The survey focuses on the issues of partner selection, intra-project
collaboration, and output performance of EU projects on the level
of bilateral partnerships, including individuals as well as organizations.
As the survey was restricted to small collaborative projects (specifically, projects
with a minimum of two and a maximum of 20 partners), the survey
addresses a subset of 9,107 relevant (59\% of all FP5) projects. It
yielded 1,686 valid responses, representing 3\% of all (relevant)
participants, and covering 1,089 (12\% of all relevant) projects.

\section{Network Definition}  \label{sec:defnetworks}

Using the sysres EUPRO database, for each FP we construct a network
containing the collaborative projects and all organizational
subentities that are participants in those projects.  An organization
is linked to a project if and only if the organization is a member
of the project.  Since an edge never exists between two organizations
or two projects, the network is bipartite.  The network edges are
unweighted; in principle, the edges could be assigned weights to
reflect the strength of the participation, but the  data needed  to
assign the network weights is not available.

We will also consider, for each FP, the projections of the bipartite
networks onto unipartite networks of organizations and projects.
The organization projections are constructed by taking the organizations
as the vertices, with edges between any organizations that are at
distance two in the corresponding bipartite network. Thus, organizations
are neighbors in the projection network if they take part in one
or more projects together. The project projections are similar,
with projects vertices linked when they have one or more participants
in common. While the construction of the projection networks intrinsically 
loses information available in the bipartite networks, they can nonetheless be 
useful.

For the binary choice model, we construct another network using cross-section 
data on 191 organizations
that are selected from the survey data. We
employ the collaboration network of the respondents on the organization
level (this network comprises 1,173 organizations collaborating in
1,089 projects) and extract the 2-core \cite{deNMrvBat:2004}
of its largest component (203 organizations representing 17\% of
all vertices)\footnote{This technical trick ensures optimal utilization
of observed collaborations in the estimation model, while keeping
the size of the model small. It is important to note that it does
not make use of the network properties on this somewhat
arbitrary sub-network.}. Finally, another 12 organizations are
excluded due to non-availability of geographical distance data, so
that we end up with a sample of 191 organizations.

\section{Network Structure} \label{sec:netstructure}

\begin{figure}
	\includegraphics[width=\textwidth]{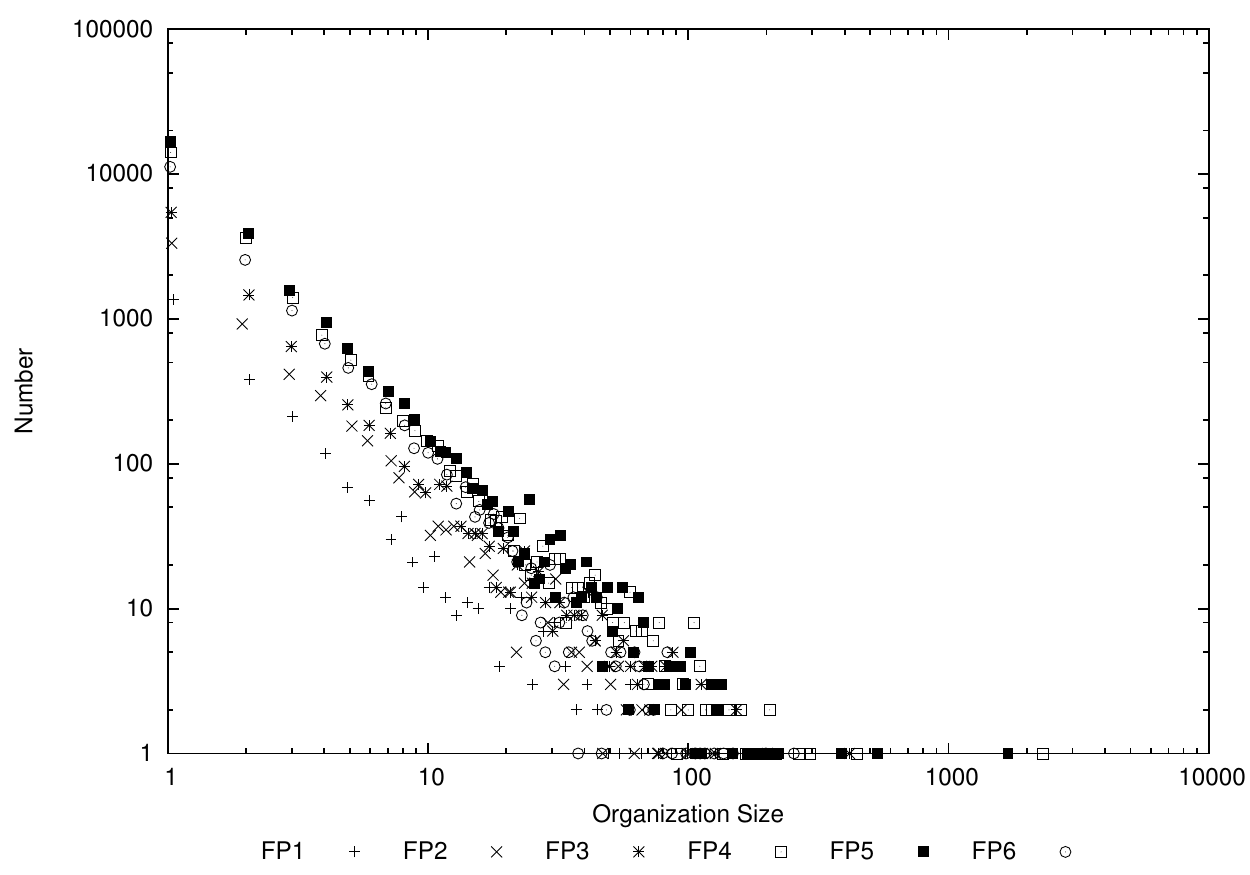}
	\caption{Organization sizes.}
	\label{fig:orgdegdist}
\end{figure}

\begin{figure}
	\includegraphics[width=\textwidth]{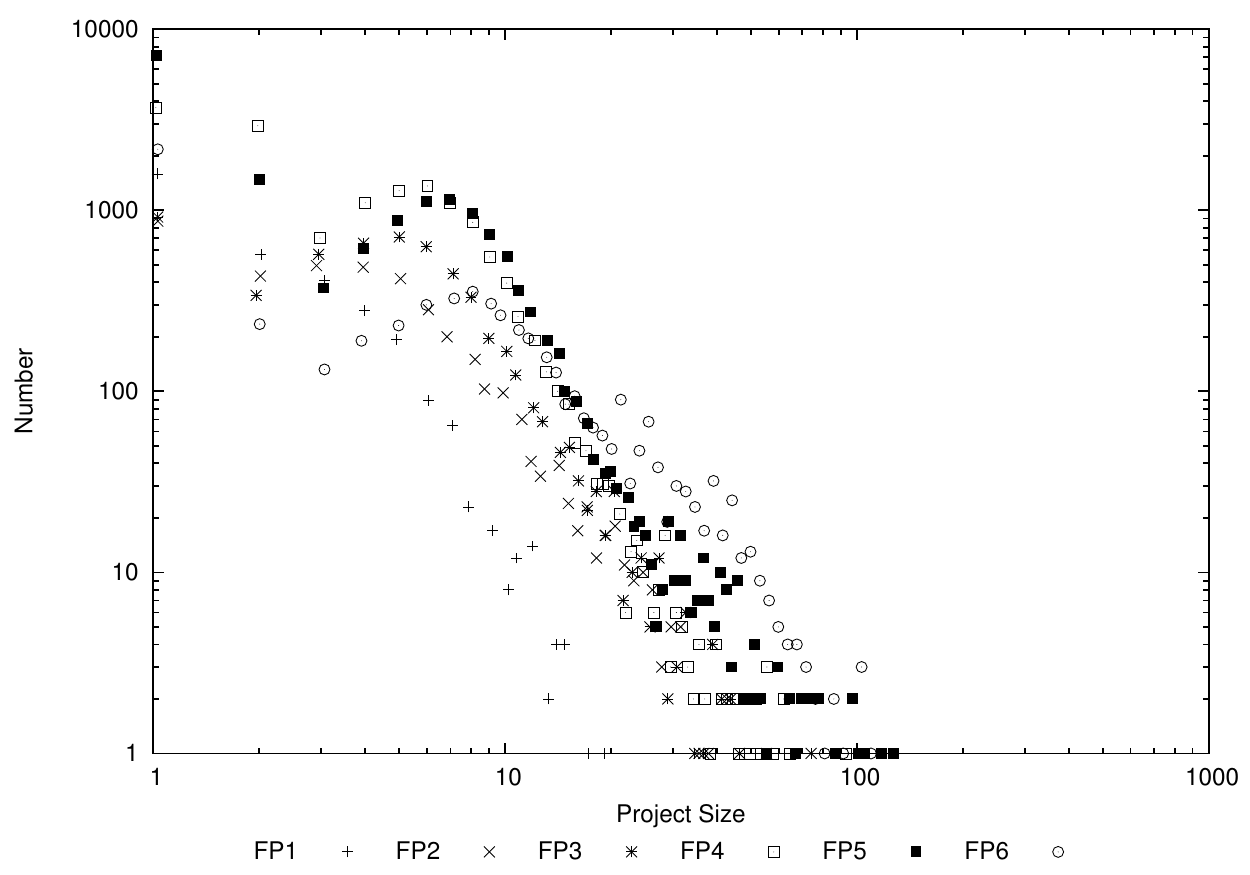}
	\caption{Project sizes.}
	\label{fig:projdegdist}
\end{figure}

We first consider the bipartite networks for each of the FP networks. Call the 
size of an organization the number of projects in which it takes part, and similarly call 
the size of a project the number of constituent organizations taking part in the project. 
These sizes correspond directly to the degrees of the relevant vertices in the bipartite
networks. Both parts---organizations (\fig{fig:orgdegdist}) and projects
(\fig{fig:projdegdist})---of each of the networks feature strongly
skewed, heavy tailed size distributions. The sizes of vertices
can differ by orders of magnitude,  pointing towards the existence
of high degree hubs in the networks; hubs of this sort can play an important role in 
determining the network structure. 

The organization size
distributions are similar for each of the FPs. The
underlying research activities thus have not altered the mix of organizations
participating in a particular number of projects in each Framework
Program, despite changes in the nature of those research activities
over time. In contrast, the rule changes in FP6 that favor larger
project consortia are clearly seen in the project size distributions.

Turning to the projection networks, we see that both the organization projection 
(\tbl{tbl:ographprops}) and the project projection (\tbl{tbl:pgraphprops}) show 
small-world properties \cite{WatStr:1998}. First, note that the great majority of 
the \( \numvertices \) vertices 
and \( \numedges \) edges are in the largest connected component of the networks. In light of 
this, we focus on paths in only the largest component. The average path length \( l \) 
in each projection network is short, as is the diameter. However, the clustering coefficient 
\cite{WatStr:1998}, which ranges between zero and one, is high. The combination of short path 
length and high clustering is characteristic of small world networks. The small-world character is expected to be beneficial in  the FP networks, as small-world networks have been shown to encourage the spread of knowledge in model systems \cite{CowJon:2004}.

\begin{table}
	\caption{Organization projection properties.}
	\begin{tabular}{lrrrrrr}
		\hline
		\tablehead{1}{c}{Measure} & \tablehead{1}{c}{FP1} & \tablehead{1}{c}{FP2} & \tablehead{1}{c}{FP3} & \tablehead{1}{c}{FP4} & \tablehead{1}{c}{FP5} & \tablehead{1}{c}{FP6} \\ 
		\hline
		No. of vertices \( \numvertices \) & 2116 & 5758 & 9035 & 21599 & 25840 & 17632 \\ 
		No. of edges \( \numedges \) & 9489 & 62194 & 108868 & 238585 & 385740 & 392879 \\ 
		No. of components & 53 & 45 & 123 & 364 & 630 & 26 \\ 
		N for largest component & 1969 & 5631 & 8669 & 20753 & 24364 & 17542 \\ 
		Share of total (\%) & 93.05 & 97.79 & 95.95 & 96.08 & 94.29 & 99.49  \\ 
		\( \numedges \) for largest component & 9327 & 62044 & 108388 & 237632 & 384316 & 392705 \\ 
		Share of total (\%) &  98.29 & 99.76 & 99.56 & 99.60 & 99.63 & 99.96  \\ 
		\( \numvertices \) for 2nd largest component & 8 & 6 & 9 & 10 & 12 & 9 \\ 
		\( \numedges \) for 2nd largest component & 44 & 30 & 72 & 90 & 132 & 72 \\ 
		Diameter of largest component & 9 & 7 & 8 & 11 & 10 & 7 \\ 
		\( l \) largest component &   3.62 &  3.21 &  3.27 &  3.45 &  3.30 &  3.03  \\ 
		Clustering coefficient &   0.65 &  0.74 &  0.74 &  0.78 &  0.76 &  0.80  \\ 
		Mean degree &    9.0 &  21.6 &  24.1 &  22.1 &  29.9 &  44.6  \\  
		Fraction of \( \numvertices \) above the mean (\%) &   29.4 &  28.0 &  23.6 &  22.4 &  23.5 &  26.1  \\ 
		\hline
	\end{tabular}
	\label{tbl:ographprops}
\end{table}

\begin{table}
	\caption{Project projection properties.}
	\begin{tabular}{lrrrrrr}
		\hline
		\tablehead{1}{c}{Measure} & \tablehead{1}{c}{FP1} & \tablehead{1}{c}{FP2} & \tablehead{1}{c}{FP3} & \tablehead{1}{c}{FP4} & \tablehead{1}{c}{FP5} & \tablehead{1}{c}{FP6} \\ 
		\hline
		No. of vertices \( \numvertices \) & 2116 & 5758 & 9035 & 21599 & 25840 & 17632 \\ 
		No. of edges \( \numedges \) & 9489 & 62194 & 108868 & 238585 & 385740 & 392879 \\ 
		No. of components & 53 & 45 & 123 & 364 & 630 & 26 \\ 
		\( \numvertices \) for largest component & 1969 & 5631 & 8669 & 20753 & 24364 & 17542 \\ 
		Share of total (\%) &  93.05 & 97.79 & 95.95 & 96.08 & 94.29 & 99.49 \\ 
		\( \numedges \) for largest component & 9327 & 62044 & 108388 & 237632 & 384316 & 392705 \\ 
		Share of total (\%) & 98.29 & 99.76 & 99.56 & 99.60 & 99.63 & 99.96 \\ 
		\( \numvertices \) for 2nd largest component & 8 & 6 & 9 & 10 & 12 & 9 \\ 
		\( \numedges \) for 2nd largest component & 44 & 30 & 72 & 90 & 132 & 72 \\ 
		Diameter of largest component & 9 & 7 & 8 & 11 & 10 & 7 \\ 
		\( l \) largest component &   3.62 &  3.21 &  3.27 &  3.45 &  3.30 &  3.03  \\ 
		Clustering coefficient &   0.65 &  0.74 &  0.74 &  0.78 &  0.76 &  0.80 \\ 
		Mean degree &    9.0 &  21.6 &  24.1 &  22.1 &  29.9 &  44.6 \\ 
		Fraction of \( \numvertices \) above the mean (\%) &   29.4 &  28.0 &  23.6 &  22.4 &  23.5 &  26.1 \\ 
		\hline
	\end{tabular}
	\label{tbl:pgraphprops}
\end{table}

Additionally, the heavy tailed size distributions of the bipartite networks has a visible effect on 
the degrees of the projection networks. In each case, the data is quite asymmetric about the 
mean degree, as seen by examining what fraction of vertices have degree above the mean. The 
fractions are between 20\% and 30\%, consistent with the skewed degree distributions (the 
distributions are shown in references~\cite{BarKruKruRoe:2006,RoeBar:2008}; the relation 
between the degrees in the bipartite networks and the projections is explored in 
reference~\cite{BarKruKruRoe:2006}).

\section{Community Structure} \label{sec:commstruct}

Of great current interest is the identification of community groups,
or modules, within networks.  Stated informally, a community group
is a portion of the network whose members are more tightly linked
to one another than to other members of the network. A variety of
approaches
\citep{AngBocMarPelStr:2007,GolKog:2006,Has:2006,NewLei:2007,ReiBor:2006,PalDerFarVic:2005,NewGir:2004,ClaNewMoo:2004,GirNew:2002}
have been taken to explore this concept; see
references~\cite{DanDiaDucAre:2005,New:2004b} for useful reviews.
Detecting community groups allows quantitative investigation of
relevant subnetworks. Properties of the subnetworks may differ from
the aggregate properties of the network as a whole, \eg,  modules
in the World Wide Web are sets of topically related web pages.
Thus, identification of community groups within a network is a first
step towards understanding the heterogeneous substructures of the
network.

Methods for identifying community groups can be specialized to
distinct classes of networks, such as bipartite networks
\cite{Bar:2007,GuiSalAma:2007}. This is immediately relevant for our study of the 
FP networks, allowing us to examine the community structure in the bipartite networks. Communities are expected to be formed of groups of organizations engaged in \RD into similar topics, and the projects in which those organizations take part. 

\subsection{Modularity}

To identify communities, we take as our starting point the
\newterm{modularity}, introduced by \citet{NewGir:2004}. Modularity
makes  intuitive notions of community groups precise by comparing
network edges to those of a null model. The modularity~\( \modularity
\) is proportional to the difference between the number of edges
within communities \( c \) and those for a null model:
\begin{equation}
	\modularity\equiv \frac{1}{2\numedges }\sum_{c}\sum_{i,j\in c}\left(
	\adjelem{i}{j}-\probelem{i}{j}\right) \mathperiod
	\label{eq:modularity}
\end{equation}
Along with \eqn{eq:modularity}, it is necessary to provide a null model, 
defining \( \probelem{i}{j} \). 

The standard choice for the null model 
constrains the degree distribution for the vertices to match the degree 
distribution in the actual network. Random graph models of this sort 
are obtained \citep{ChuLu:2002} by putting an edge between 
vertices \( i \) and \( j \) at random, with
the constraint that on average the degree of any vertex \( i \) 
is \( \vertexdegree{i} \). This constrains the expected adjacency matrix such that 
\begin{equation}
	\vertexdegree{i}=E\left( \sum_{j}\adjelem{i}{j}\right) \mathperiod
\end{equation}
Denote \( E\left( \adjelem{i}{j}\right)  \) by \( \probelem{i}{j} \) and assume further that \( \probelem{i}{j} \)
factorizes into
\begin{equation}
	\probelem{i}{j}=p_{i}p_{j} \mathcomma
\end{equation}
leading to
\begin{equation}
	\probelem{i}{j}\equiv \frac{\vertexdegree{i}\vertexdegree{j}}{2\numedges } \mathperiod
\end{equation}
A consequence of the null model choice is that \( \modularity = 0 \) when all vertices are in 
the same community.

The goal now is to find a division of the vertices into communities such
that the modularity \( \modularity \) is maximal. An exhaustive search for a decomposition
is out of the question: even for moderately large graphs there are far too
many ways to decompose them into communities. Fast approximate algorithms do
exist (see, for example, references~\citep{PujBejDel:2006,New:2004a}). 

\subsection{Finding Communities in Bipartite Networks}

Specific classes of networks have additional constraints that can be reflected in the null model.
For bipartite graphs, the null model should be modified to reproduce the
characteristic form of bipartite adjacency matrices:
\begin{equation}
	\adjmat=\bipartsubstructure{\adjsubmat}
	\mathperiod
	\label{eq:bipartsubstructure}
\end{equation}
Recently, specialized modularity measures and search algorithms have been proposed for
finding communities in bipartite networks \citep{Bar:2007,GuiSalAma:2007}. These measures
and methods have not been studied as extensively as the versions with the standard null model shown above, but many of the algorithms can be adapted to the bipartite versions without difficulty. Limitations of modularity-based methods (\eg, the resolution limit described in reference~\cite{ForBar:2007}) are expected to hold as well.  

We make use of the algorithm called \brim: bipartite, recursively induced 
modules \citep{Bar:2007}. \brim is a conceptually simple, greedy search algorithm that 
capitalizes on the separation between the two parts of a bipartite network.
Starting from some partition of the vertices of type 1, it
is straightforward to identify the optimal partition of the vertices
of type 2. From there, optimize the partition of vertices of type 1, and
so on. In this fashion, modularity  increases until a (local) maximum
is reached. However, the question remains: is the maximum a ``good'' one? At this 
level then a random search is called
for, varying the composition and number of communities, with
the goal of reaching a better maximum after a new sequence of searching using 
the \brim algorithm.

\section{Communities in the Framework Program Networks} \label{sec:fpcommunities}

A popular approach in social network
analysis---where networks are often small, consisting of a few dozen
nodes---is to visualize the networks and identify community groups
by eye. However, the Framework Program networks are much larger:
can we ``see'' the community groups in these networks?

Structural differences or similarities of such networks are not obvious at a
glance. For a graphical representation of the organizations and/or projects
by dots on an A4 sheet of paper, we would need to put these dots at a
distance of about \( 1\,\mathrm{mm} \) 
 from each other, and we then still would not have drawn the
links (collaborations) which connect them.

Previous studies used a list of coarse graining recipes to compact the networks into a form which would lend itself to a graphical representation \cite{BreCus:2004}.
As an alternative we have attempted to detect communities just using \brim,
\ie, purely on the basis of relational network structure, ignoring
any additional information about the nature of agents.

\begin{figure}
	\includegraphics[width=\textwidth]{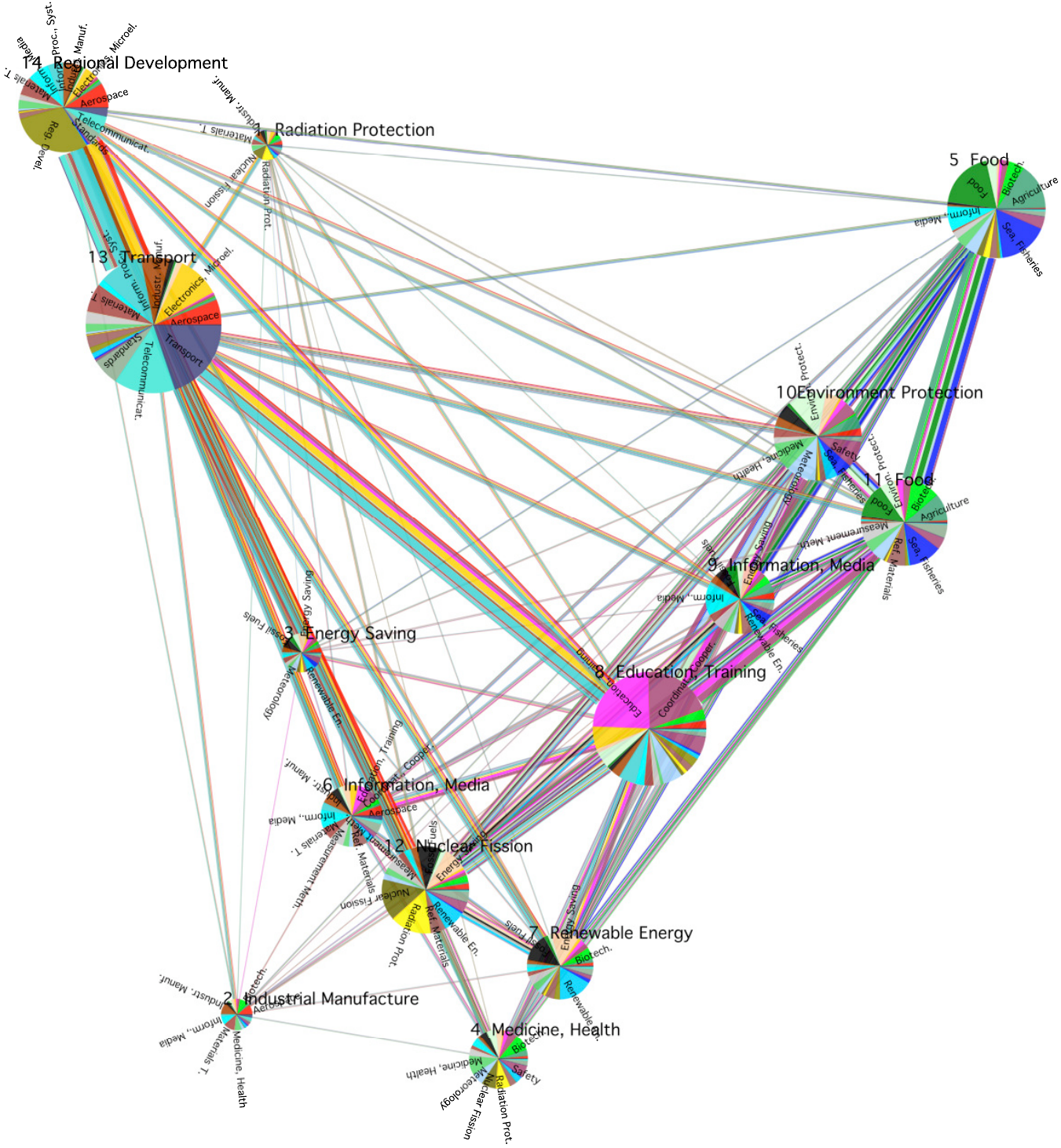}
	\caption{Community groups in the network of projects and organizations for FP3.}
	\label{fig:brimclusters}
\end{figure}

In \fig{fig:brimclusters}, we show a community structure for FP3
found using the \brim algorithm, with a modularity of \( \modularity
= 0.602 \) for 14 community groups. The communities are shown as
vertices in a network, with the vertex positions determined using
spectral methods \citep{SeaRic:2003}.  The area of each vertex is
proportional to the number of edges from the original network within
the corresponding community.  The width of each edge in the community
network is proportional to the number of edges in the original
network connecting community members from the two linked groups.
The vertices and edges are shaded to provide additional information
about their topical structure, as described in the next section.
Each community is labeled with the most frequently occurring subject
index.

\subsection{Topical Profiles of Communities}

Projects are assigned one or more standardized subject indices. There are 49 subject indices in total, 
ranging from \subjectindex{Aerospace} to \subjectindex{Waste Management}.
We denote by
\begin{equation}
	\freq{t} > 0 
\end{equation}
the frequency of occurrence of the subject index \( t \) in the network, with 
\begin{equation}
	\sum\limits_{t}\freq{t} = 1\mathperiod
\end{equation}
Similarly we consider the projects within one community \( c \) and the frequency 
\begin{equation}
	\topfreq{c}{t}\geq 0 
\end{equation}
of any subject index \(  t \) appearing in the projects only of that community. We
call \( \topfreqnoarg{c} \) the topical profile of community \( c \) to be compared with that of
the network as a whole. 

Topical differentiation of communities can be measured by comparing their
profiles, among each other or with respect to the overall network. This can be done
in a variety of ways \citep{GibSu:2002}, such as by the Kullback ``distance''
\begin{equation}
	D_{c}=\sum\limits_{t} \topfreq{c}{t}\ln \frac{\topfreq{c}{t}}{\freq{t}} \mathperiod
\end{equation}
A true metric is given by
\begin{equation}
	\topicmetric{c} = \sum\limits_{t} \abs{ \topfreq{c}{t}-\freq{t} } \mathcomma
\end{equation}
ranging from zero to two.

\begin{figure}
	\includegraphics[width=\textwidth]{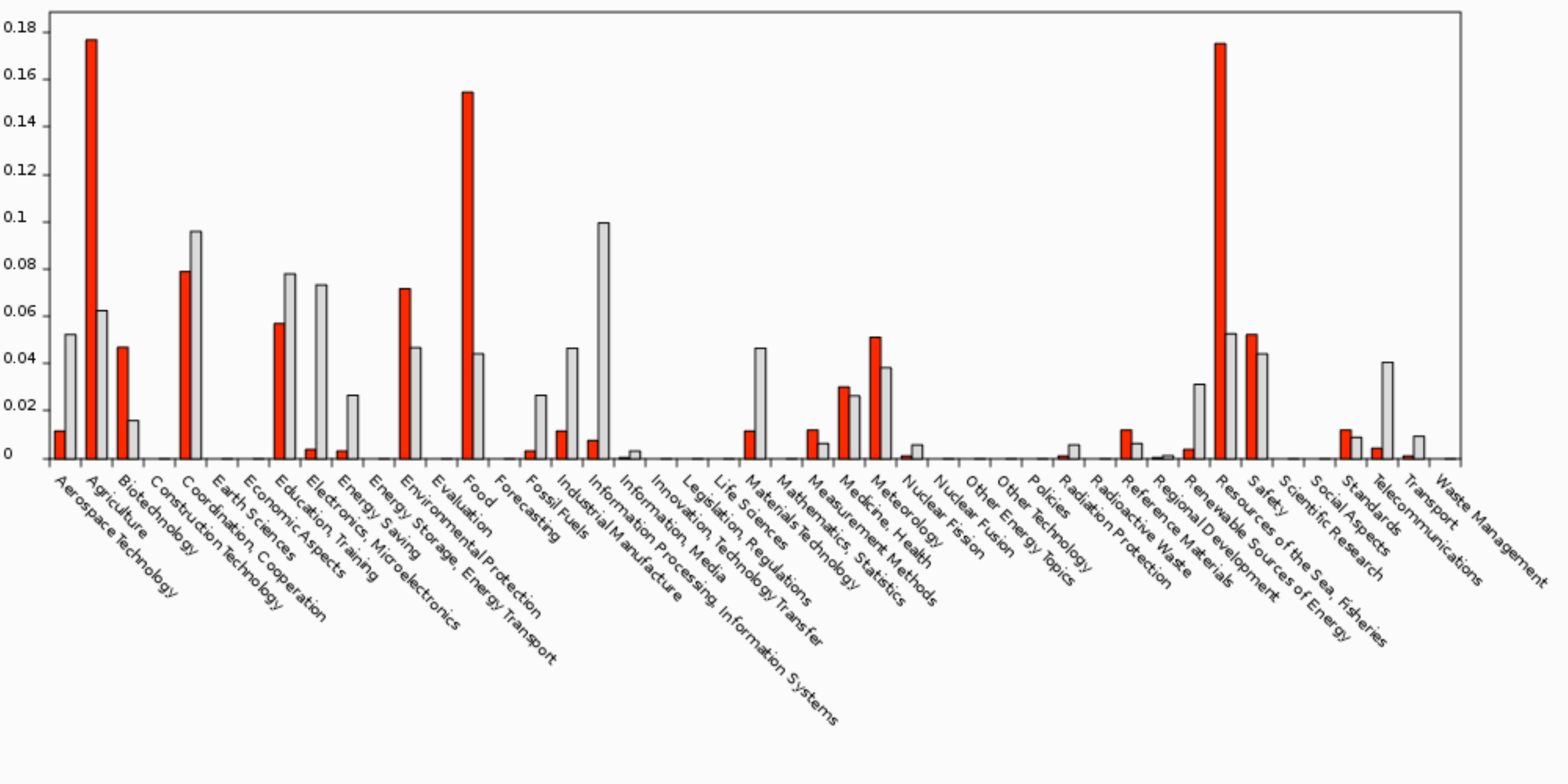}
	\caption{Topical differentiation in a network community.
	The histogram shows the difference between the topical
	profile \( \topfreq{c}{t} \) for a specific community (dark
	bars) and the overall profile \( \freq{t} \) for the network
	as a whole (light bars). The community-specific profile
	shown is for the community labeled ``11.~Food'' in
	\fig{fig:brimclusters}. The community has \( \topicmetric{c}
	= 0.90 \). }
	\label{fig:topichistogram}
\end{figure}

Topical differentiation is illustrated in \fig{fig:topichistogram}.
In the figure, example profiles are shown, taken from the network
in \fig{fig:brimclusters}. The community-specific profile corresponds
to the community  labeled `11.~Food''  in \fig{fig:brimclusters}.
Based on the most frequently occurring subject
indices---\subjectindex{Agriculture}, \subjectindex{Food}, and
\subjectindex{Resources of the Seas, Fisheries}---the community
consists of projects and organizations focussed on \RD related to
food products. The topical differentiation is \( \topicmetric{c} =
0.90 \) for the community shown.

\section{Binary Choice Model} \label{sec:binchoicemodel}

We now turn to modeling  organizational
collaboration choices in order to examine how specific
individual characteristics, spatial effects, and network effects
determine the choice of collaboration (the theoretical underpinnings are described in 
reference~\cite{PaiSch:2008}. We will build upon the 
survey data and the subnetwork constructed therefrom (\sxn{sec:defnetworks}).  
While this restricts us to only 191 organizations, we have considerably more information
about these organizations than for the complete networks.

\subsection{The Empirical Model}

In our analytical framework,
the constitution of a collaboration \( \collabelem{i}{j} \) between two organizations
\( i \) and \( j \) will depend on an unobserved continuous variable \( \hiddenprofitelem{i}{j} \) that corresponds
to the profit that two organizations \( i \) and \( j \) receive when they
collaborate. Since we cannot observe \( \hiddenprofitelem{i}{j} \) but only its dichotomous
realizations \( \collabelem{i}{j} \), we assume \( \collabelem{i}{j} = 1 \) if \( \hiddenprofitelem{i}{j} > 0\) and \( \collabelem{i}{j} = 0 \) if  \( \hiddenprofitelem{i}{j} \leq 0\). \( \collabelem{i}{j} \) is assumed
to follow a Bernoulli distribution so that \( \collabelem{i}{j} \) can take the values one
and zero with probabilities \( \oneprob{i}{j} \) and  \(\zeroprob{i}{j} \),
respectively. The probability function can be written as
\begin{equation}
	\probfunc{\collabelem{i}{j}} = \oneprob{i}{j}^{\collabelem{i}{j}} \largeparens{\zeroprob{i}{j}}^{1-\collabelem{i}{j}} 
\end{equation}
with \( \expectationop{ \collabelem{i}{j} } = \collabmeanelem{i}{j} = \oneprob{i}{j} \) and  \( \varianceop{ \collabelem{i}{j} } = \collabvarelem{i}{j} = \oneprob{i}{j}\largeparens{\zeroprob{i}{j}} \), where \( \collabmeanelem{i}{j} \) denotes some mean value. 

The next step in defining the model  concerns
the systematic structure---we would like the probabilities \( \oneprob{i}{j} \) to
depend on a matrix of observed covariates. Thus, we let the probabilities \( \oneprob{i}{j} \)
be a linear function of the covariates:
\begin{equation}
	\oneprob{i}{j} = \sum_{k=1}^{\numexpvars} \paramwtelem{k} \covariateelem{k}{i}{j}
	\mathcomma
	\label{eq:lincovfun}
\end{equation}
where the \( \covariateelem{k}{i}{j} \) are elements of the \( \covariatemat{k} \) matrix containing a constant and \( \numexpvars - 1 \) explanatory
variables, including geographical effects, relational effects and
FP experience characteristics of organizations \( i \) and \( j \). \( \paramwtvctr_{\numexpvars} = \litvctr{ \paramwtelem{0}, \paramwtvctr_{\numexpvars-1} } \) is the
\( \matrixdims{\numexpvars}{1} \) parameter vector, where \( \paramwtelem{0} \) is a 
scalar constant term and \( \paramwtvctr_{\numexpvars-1} \) is
the vector of parameters associated with the \( \numexpvars - 1 \) explanatory
variables.

However, estimating this model using ordinary least squares procedures is not convenient
since the probability \( \oneprob{i}{j} \) must be between zero and one, while the
linear predictor can take any real value. Thus, there is no guarantee
that the predicted values will be in the correct range without
imposing any complex restrictions\cite{JohDin:2007}.
A very promising solution to this problem is to use the logit
transform of \( \oneprob{i}{j} \) in the model, \ie, replacing \eqn{eq:lincovfun} by 
the following \foreign{ansatz}:
\begin{equation}
	\logit{\oneprob{i}{j}} = \log \frac{\oneprob{i}{j}}{\zeroprob{i}{j}} = 
	 \covwtsumelem{i}{j} 
	 \mathcomma
	 \label{eq:logitansatz}
\end{equation}
where we have introduced the abbreviation \( \covwtsumelem{i}{j} \), defined as
\begin{equation}
	\covwtsumelem{i}{j} = \paramwtelem{0} + \paramwtelem{1}\covariateelem{1}{i}{j} + \paramwtelem{2}\covariateelem{2}{i}{j} + \cdots + \paramwtelem{\numexpvars}
	\mathperiod
\end{equation}
This leads to the binary logistic regression model to be estimated given by
\begin{equation}
	\probfunc{\collabelem{i}{j}=1 \given \covariateelem{k}{i}{j}} = \oneprob{i}{j} = 
	\frac{\exp\largeparens{ \covwtsumelem{i}{j} }} {1+\exp\largeparens{ \covwtsumelem{i}{j} }}
	\mathperiod
	\label{eq:binlogregmodel}
\end{equation}
The focus of interest is on estimating the parameters \( \paramwtvctr \). The standard
estimator for the logistic model is the maximum likelihood estimator.
The reduced log-likelihood function is given by \cite{JohDin:2007}
\begin{equation}
	\loglikelihood{ \paramwtvctr \given \collabelem{i}{j} } = 
		-\sum_{i,j} \log \largeparens{ 1+\exp\largeparens{ \largeparens{ 1- 2\collabelem{i}{j} } \covwtsumelem{i}{j} }}
		\mathcomma
\end{equation}
assuming independence over the observations \( \collabelem{i}{j} \). The resulting
variance matrix \( \variancematop{\paramestvctr} \) of the parameters is used to 
calculate standard
errors. \( {\paramestvctr} \) is consistent and asymptotically efficient when the
observations of \( \collabelem{i}{j} \) are stochastic and in absence of multicollinearity
among the covariates.

\subsection{Variable Construction}

\subsubsection{The Dependent Variable}

To construct the dependent variable \( \collabelem{i}{j} \) that corresponds to observed
collaborations between two organizations \( i \) and \( j \), we construct the
\( \matrixdims{n}{n} \) collaboration matrix \( \collabmat \) that contains the collaborative links
between the \orgpair{i}{j}-organizations. One element \( \collabelem{i}{j} \) denotes the
existence of collaboration between two organizations \( i \) and \( j \) as
measured in terms of the existence of a common project. \( \collabmat \) is symmetric
by construction so that \( \collabelem{i}{j} = \collabelem{j}{i} \). Note that the matrix is very
sparse. The number of observed collaborations is 702 so that
proportion of zeros is about 98\%. The mean collaboration intensity
between all \orgpair{i}{j}-organizations is 0.02.

\subsubsection{Variables Accounting for Geographical Effects}

We use two variables \( \obscovelem{1}{i}{j} \) and \( \obscovelem{2}{i}{j} \) to account for geographical effects on
the collaboration choice. The first step is to assign specific
NUTS-2 regions to each of the 191 organizations that
are given in the sysres EUPRO database.
Then we take the great circle distance between the economic centers
of the regions where the organizations \( i \) and \( j \) are located to measure
the geographical distance variable \( \obscovelem{1}{i}{j} \). The second variable, \( \obscovelem{2}{i}{j} \), controls
for country border effects and is measured in terms of a dummy
variable that takes a value of zero if two organizations \( i \) and \( j \)
are located in the same country, and zero otherwise, in order to
get empirical insight on the role of country borders for collaboration
choice of organizations.

\subsubsection{Variables Accounting for FP Experience of Organizations}

This set of variables controls for the experience of the organizations
with respect to participation in the European FPs. First, thematic
specialization within FP5 is expected to influence the potential
to collaborate. We define a measure of thematic distance \( \obscovelem{3}{i}{j} \) between
any two organizations that is constructed in the following way:
Each organization is associated with a unit vector of specialization \( \specvctr{i} \)
that relates to the number of project participations \( \numprojpart{i}{1}, \ldots, \numprojpart{i}{7} \) of organization
\( i \) in the seven sub-programs of FP5\footnote{EESD, GROWTH, 
HUMAN POTENTIAL, INCO 2, INNOVATION-SME, IST, and LIFE 
QUALITY}.
\begin{equation}
	 \specvctr{i} = \litvctr{\numprojpart{i}{1}, \ldots, \numprojpart{i}{7}} / 
	 	\sqrt{\numprojpart{i}{1} ^ 2 + \cdots + \numprojpart{i}{7} ^ 2}
	 	\mathperiod
\end{equation}
The thematic distance of organizations \( i \) and \( j \) is then defined as
the Euclidean distance of their respective specialization vectors \( \specvctr{i} \)
and \( \specvctr{j} \),
giving \( \obscovelem{3}{i}{j} = \obscovelem{3}{j}{i} \) and \( 0 \leq \obscovelem{3}{i}{j} \leq \sqrt{2} \). The second variable accounting for
FP experience focuses on the individual (or research group) level,
and takes into account the respondents inclination or openness to
FP research. As a proxy for openness of an organization \( i \) to FP
research, we choose the total number \( \numprojknown{i} \) of FP5 projects in the
respondent's own organization, that they are aware of\footnote{The
exact wording of the question was, `How many FP5 projects of your
organization are you aware of?' For multiple responses from an
organization, the numbers of known projects are summarized. In cases
of missing data, this number is set to zero. }. Then we define
\begin{equation}
	\obscovelem{4}{i}{j} = \numprojknown{i} + \numprojknown{j}
\end{equation}
as a measure for the aggregated openness of the respective pair of
organizations to FP research. The third variable related with FP
experience is the overall number of FP5 project participations an
organization is engaged in. Denoting, as above, \( \totprojpart{i} = \numprojpart{i}{1} + \cdots + \numprojpart{i}{7} \) as the total
number of project participations of organization \( i \) in FP5, we define
\begin{equation}
	\obscovelem{5}{i}{j} = \abs{\totprojpart{i} - \totprojpart{j}}
\end{equation}
as the difference in the number of participations of organization
\( i \) and \( j \) in FP5. It is taken from the sysres EUPRO database and is
an integer ranging from \( 0 \leq \obscovelem{5}{i}{j} \leq  \) 1,228, resulting from the minimal value of one
participation and the maximum of 1,229 participations among the
sample of 191 organizations.

\subsubsection{Variables Accounting for Relational Effects} 
We consider a set of three variables accounting for potential
relational effects on the decision to collaborate. Hereby, we
distinguish between joint history and network effects. The first
factor to be taken into account is prior acquaintance of two
organizations, and is measured by a binary variable denoting
acquaintance on the individual (research group) level before the
FP5 collaboration started. It is taken from the survey\footnote{The exact wording of the question was, `Which of your [project acronym] partners (\ie, persons from which organization) did you know before the project began?'}. By convention, \( \obscovelem{6}{i}{j} = 1 \) if at least one respondent from organization \( i \) nominated organization \( j \)
as prior acquainted, \( \obscovelem{5}{i}{j} = 0 \) otherwise. All other relational factors we
take into account in the model are network effects. For conceptual
reasons we must look at the global FP5 network, where we make
use of the structural embeddedness of our 191 sample organizations.

One of the most important centrality measures is betweenness
centrality. Betweenness is a centrality concept based on the question
to what extent a vertex in a network is able to control the information
flow through the whole network \cite{WasFau:1994}. Organizations
that are high in betweenness, may thus be especially attractive as
collaboration partners. More formally, the betweenness centrality
of a vertex can be defined as the probability that a shortest path
between a pair of vertices of the network passes through this vertex.
Thus, if \( \numshortestbetthru{k}{l}{i} \) is the number of shortest paths between vertices \( k \) and \( l \)
passing through vertex \( i \), and \( \numshortestbet{k}{l} \) is the total number of shortest paths
between vertices \( k \) and \( l \), then
\begin{equation}
	\betcent{i} = \sum_{k  \neq l} \frac{\numshortestbetthru{k}{l}{i}}{\numshortestbet{k}{l}}
\end{equation}
is called the betweenness centrality of vertex \( i \) \cite{DorMen:2004}. We calculate the betweenness centralities in the
global FP5 network and include
\begin{equation}
	 \obscovelem{7}{i}{j} = \betcent{i} \betcent{j}
\end{equation}
as a combined betweenness measure. 

The third variable accounting
for relational effects is local clustering. Due to social closure,
we may assume that within densely connected clusters organizations
are mutually quite similar, so that it might be strategically
advantageous to search for complementary partners from outside.
Hereby, communities with lower clustering may be easier to access.
We use the clustering coefficient \( \locclustcoef{i} \), which is the share of existing
links in the number of all possible links in the direct neighborhood
(at distance \( d=1 \)) of a vertex \( i \). Thus, let \( \degree{i} \) be the number of direct
neighbors and \( \numtris{i} \) the number of existing links among these direct
neighbors, then
\begin{equation}
	\locclustcoef{i} = \frac{ 2 \numtris{i} }{ \degree{i} \largeparens{\degree{i}-1} }
\end{equation}
is the relevant clustering coefficient \cite{WatStr:1998}.
We employ the difference in the local clustering coefficients within
the global FP5 network for inclusion in the statistical model, by
setting
\begin{equation}
	 \obscovelem{8}{i}{j} = \abs{\locclustcoef{i} - \locclustcoef{j}}
\end{equation}
in order to obtain a symmetric variable in \( i \) and \( j \).

\subsection{Estimation Results}

This section discusses the estimation results of the binary choice
model of \RD collaborations as given by \eqn{eq:binlogregmodel}. The
binary dependent variable corresponds to observed collaborations
between two organizations \( i \) and \( j \), taking a value of one if they
collaborate and zero otherwise. The independent variables are
geographical separation variables, variables capturing FP experience
of the organizations and relational effects (joint history and
network effects). We estimate three model versions: The standard
model includes one variable for geographical effects and FP experience,
respectively, and two variables accounting for relational effects.
In the extended model version we add country border effects as
additional geographical variable in order to isolate country border
effects from geographical distance effects, and openness to FP
research as additional FP experience variable. The full model
additionally includes balance variables accounting for FP experience
and network effects, respectively.

\Tbl{tbl:estresults} presents the sample estimates derived from maximum likelihood
estimation for the model versions. The number of observations is
equal to 36,481, asymptotic standard errors are given in parentheses.
The statistics given in \tbl{tbl:modelperformance} indicate that the
selected covariates show a quite high predictive ability. The
Goodman-Kruskal-Gamma statistic ranges from 0.769 for the basic and
0.782 for the extended model to 0.786 for the full model, indicating
that more than 75\% fewer errors are made in predicting interorganizational
collaboration choices by using the estimated probabilities than by
the probability distribution of the dependent variable alone. The
Somers \( D \) statistic and the \( C \) index confirm these findings. The
Nagelkerke's \( R \)-Squared is 0.391 for the basic model, 0.395 for the
extended model and 0.397 for the full model version, 
respectively\footnote{Nagelkerke's R-squared is an attempt to imitate 
the interpretation of multiple R-Squared measures from linear regressions
 based on the log likelihood of the final model versus log likelihood of the 
 null model. It is defined as 
 \( R_{\mathrm{Nag}}^2 = \largesquare{ 1 = \largeparens{L_0 / L_1 }^{2/n}} / \largesquare{ 1 - L_0^{2/n}} \)
 where \( L_0 \) is the log likelihood of the null model, \( L_1 \) is the log 
 likelihood of the model to be evaluated and \( n \) is the number of 
 observations.}.
A likelihood ratio test for the null hypothesis of \( \paramwtelem{k} = 0 \) yields a \( \chi_4^2 \) test
statistic of 2,565.165 for the basic model, a \( \chi_6^2 \) test statistic of
2,582.421 for the extended model and a \( \chi_8^2 \) test statistic of  2,597.911
for the full model. These are statistically significant and we
reject the null hypothesis that the model parameters are zero for
all model versions.

\begin{table}
	\caption{Maximum likelihood estimation results for the collaboration model based on \(n^2\)=36,481 observations. Asymptotic standard errors are given parenthetically.}
	\begin{tabular}{lrrr}
		\hline
		\tablehead{1}{c}{Coefficient} & \tablehead{1}{c}{Basic Model} & \tablehead{1}{c}{Extended Model} & \tablehead{1}{c}{Full Model} \\ 
		\hline
		\( \paramwtelem{0} \) & -1.882\vhighsig(0.313) & -1.951\vhighsig(0.342) & -1.816\vhighsig(0.385) \\ 
		\( \paramwtelem{1} \) & -0.145\vhighsig(0.038) & -0.116\vhighsig(0.039) & -0.128\vhighsig(0.040) \\ 
		\( \paramwtelem{2} \) & \missingtableentry & -0.103\vhighsig(0.034) & -0.094\highsig(0.034) \\ 
		\( \paramwtelem{3} \) & -1.477\vhighsig(0.110) & -1.465\vhighsig(0.114) & -1.589\vhighsig(0.117) \\ 
		\( \paramwtelem{4} \) & \missingtableentry & 0.004\vhighsig(0.001) & 0.003\vhighsig(0.001) \\ 
		\( \paramwtelem{5} \) & \missingtableentry & \missingtableentry & 0.001(0.000) \\ 
		\( \paramwtelem{6} \) & 4.224\vhighsig(0.089) & 4.189\vhighsig(0.089) & 4.194\vhighsig(0.089) \\ 
		\( \paramwtelem{7} \) & 0.161\vhighsig(0.023) & 0.135\vhighsig(0.025) & 0.119\vhighsig(0.027 \\ 
		\( \paramwtelem{8} \) & \missingtableentry & \missingtableentry & 0.070\highsig(0.025) \\
		\hline
	\end{tabular}
	\label{tbl:estresults}
\end{table}

\begin{table}
	\caption{Performance of the three collaboration model versions based on \(n^2\)=36,481 observations. }
		\begin{tabular}{lrrr}
			\hline
			\tablehead{1}{c}{Performance}  & \tablehead{1}{c}{Basic Model} & \tablehead{1}{c}{Extended Model} & \tablehead{1}{c}{Full Model} \\ 
			\hline
			Somers \(D\) & 0.733 & 0.746 & 0.753 \\ 
			Goodman-Kruskal Gamma & 0.769 & 0.782 & 0.786 \\ 
			\(C\) index & 0.876 & 0.873 & 0.875 \\ 
			Nagelkerke \(R\)-squared index & 0.391 & 0.395 & 0.397 \\ 
			Log-likelihood & -2,190.151 & -2,176.768 & -2,169.578 \\ 
			Likelihood Ratio Test & 2,565.156\vhighsig & 2,582.421\vhighsig & 2,597.911\vhighsig \\
			\hline
		\end{tabular}
		\label{tbl:modelperformance}
\end{table}

The model reveals some promising empirical insight in the context
of the relevant literature on innovation as well as on social
networks. The results provide a fairly remarkable confirmation of
the role of geographical effects, FP experience effects and network
effects for interorganizational collaboration choice in EU FP \RD
networks. In general, the parameter estimates are statistically
significant and quite robust over different model versions.

The results of the basic model show that geographical distance
between two organizations significantly determines the probability
to collaborate. The parameter estimate of \( \paramwtelem{1} = -0.145 \)  indicates that
for any additional 100 km between two organizations the mean
collaboration frequency decreases by about 15.6\%. Geographical
effects matter but effects of FP experience of organizations are
more important. As evidenced by the estimate \( \paramwtelem{3} = -1.477 \) it is most
likely that organizations choose partners that are located closely
in thematic space. A one percent increase in thematic distance
reduces the probability of collaboration by more than 3.25\%.
Most important determinants of collaboration choice are network
effects. The estimate of \( \paramwtelem{6} = 4.224 \) tells us that the probability of
collaboration between two organizations increases by 68.89\%
when they are prior acquaintances. Also network embeddedness matters
as given by the estimate for \( \paramwtelem{7} = 0.161 \) indicating that  choice of
collaboration is more likely between organizations that are central
players in the network with respect to betweenness centrality.

Turning to the results of the extended model version it can be seen
that taking into account country border effects decreases geographical
distance effects by about 24\% (\( \paramwtelem{1} = -0.116 \)). The existence of a
country border between two organizations has a significant negative
effect on their collaboration probability, the effect is slightly
smaller than geographical distance effects (\( \paramwtelem{2} = -0.103 \)). Adding
openness to FPs as an additional variable to capture FP experience
does not influence the other model parameters much. Openness to FPs, though statistically 
significant, shows only a small impact on collaboration
choice.

In the full model version we add one balance variable accounting
for FP experience and network effects, respectively. The difference
in the number of submitted FP projects has virtually no effect on
the choice of collaboration as given by the estimate of \( \paramwtelem{5} \). An
interesting result from a social network analysis perspective
provides the integration of the difference between two organizations
with respect to the clustering coefficient. The
estimate of \( \paramwtelem{8} = 0.070 \) tells us that it is more likely that two
organizations collaborate when the difference of their cluster
coefficients is higher. This result points to the existence of
strategic collaboration choices for organizations that are highly
cross-linked searching for organizations to collaborate with lower
clustering coefficients, and the other way round. The effect is
statistically significant but smaller than other network effects
and geographical effects.

\section{Summary} \label{sec:summary}

We have presented an investigation of networks derived from the
European Union's Framework Programs for Research and Technological
Development. The networks are of substantial size, complexity, and 
economic importance. We have attempted to provide a coherent picture of the
complete process, beginning with data preparation and network
definition, then continuing with analysis of the network structure and modeling 
of network formation. 

We first considered the challenges involved in dealing
with a large amount of imperfect data, detailing the tradeoffs made
to clean the raw data into a usable form under finite resource
constraints.  The processed data was then used to define bipartite
networks with vertices consisting of all the projects and organizations
involved in each FP. To provide alternative views of the data, we
defined projection networks for each part (organizations or projects)
of the bipartite networks. Additionally, we used results of a survey
of FP5 participants to define a smaller network about which we have
more detailed information than we have for the networks as a whole.

Next we examined structural properties of the bipartite and projection networks. We 
found that the vertex degrees in the FP networks have a highly skewed, heavy tailed 
distribution. The networks further show characteristic features of small-world networks, 
having both high clustering coefficients and short average path lengths. 
We followed this with analysis of the community structure of the Framework Programs. 
Using a modularity measure and search algorithm adapted to bipartite networks, we 
identified communities from the networks, and found that the communities are topically
differentiated based on the standardized subject indices for Framework Program projects. 

In the final stage of analysis, we constructed a binary choice model to explore
determinants of inter-organizational collaboration choice. The model
parameters were estimated using logistic regression. The model results show
that geographical effects matter, but are not the most important determinants.
The strongest effect comes from relational characteristics, in particular prior acquaintance, 
and to a minor extent, network centrality. Also, thematic similarity between organizations 
highly favors a partnership. 

By using a variety of networks and analyses, we have been able to
address several different questions about the Framework Programs.
The results complement one another, giving a more complete picture of
the Framework Programs than the results from any one method alone.
We are confident that our understanding of collaborative \RD in the
European Union can be improved by extending the analyses presented
in this chapter and by expanding the types of analyses we undertake.

\acknowledgments

The authors gratefully acknowledge financial support from the European FP6-NEST-Adventure Program, under contract number~028875 (project NEMO: Network Models, Governance, and \RD Collaboration Networks).

%: Bibliography

\bibliographystyle{plainnat}
\bibliography{references}

\end{document}